\documentclass[prb,preprint]{revtex4-1}

\usepackage{amsmath}
\usepackage{graphicx}

\newcommand{\ang}{L} 

\begin{document}
\title{The weight of a falling chain, revisited}

\author{Eugenio Hamm}
\email{luis.hamm@usach.cl}
\affiliation{Universidad de Santiago de Chile, av. Ecuador 3493, 9170124 Estaci\'on Central, Santiago de Chile}
\affiliation{Center for Advanced Interdisciplinary Research on Materials, CIMAT Fondap, Av.\ Blanco Encalada 2008, Casilla 487-3, 837.0415, Santiago de Chile }

\author{Jean-Christophe G\'eminard}
\email{Jean-Christophe.Geminard@ens-lyon.fr}
\affiliation{Universit\'e de Lyon, \'Ecole Normale Sup\'erieure de Lyon, CNRS, Laboratoire de Physique, 46 All\'ee d'Italie, 69364 Lyon Cedex 07, France}


\begin{abstract}
A vertically hanging chain is released from rest and falls due to gravity on a scale pan.
We discuss the various experimental and theoretical aspects of this classic problem.
Careful time-resolved force measurements allow us to determine the differences between the idealized and the its implementation in the laboratory problem.
We observe that, in spite of the upward force exerted by the pan on the chain,
the free end at the top falls faster than a freely falling body.
Because a real chain exhibits a finite minimum radius of curvature,
the contact at the bottom results in a tensional force which pulls the falling part downward.
\end{abstract}

\maketitle

\section{\label{sec0}Introduction}

\begin{figure}[h!]
\includegraphics[height=0.8\columnwidth,angle=-90]{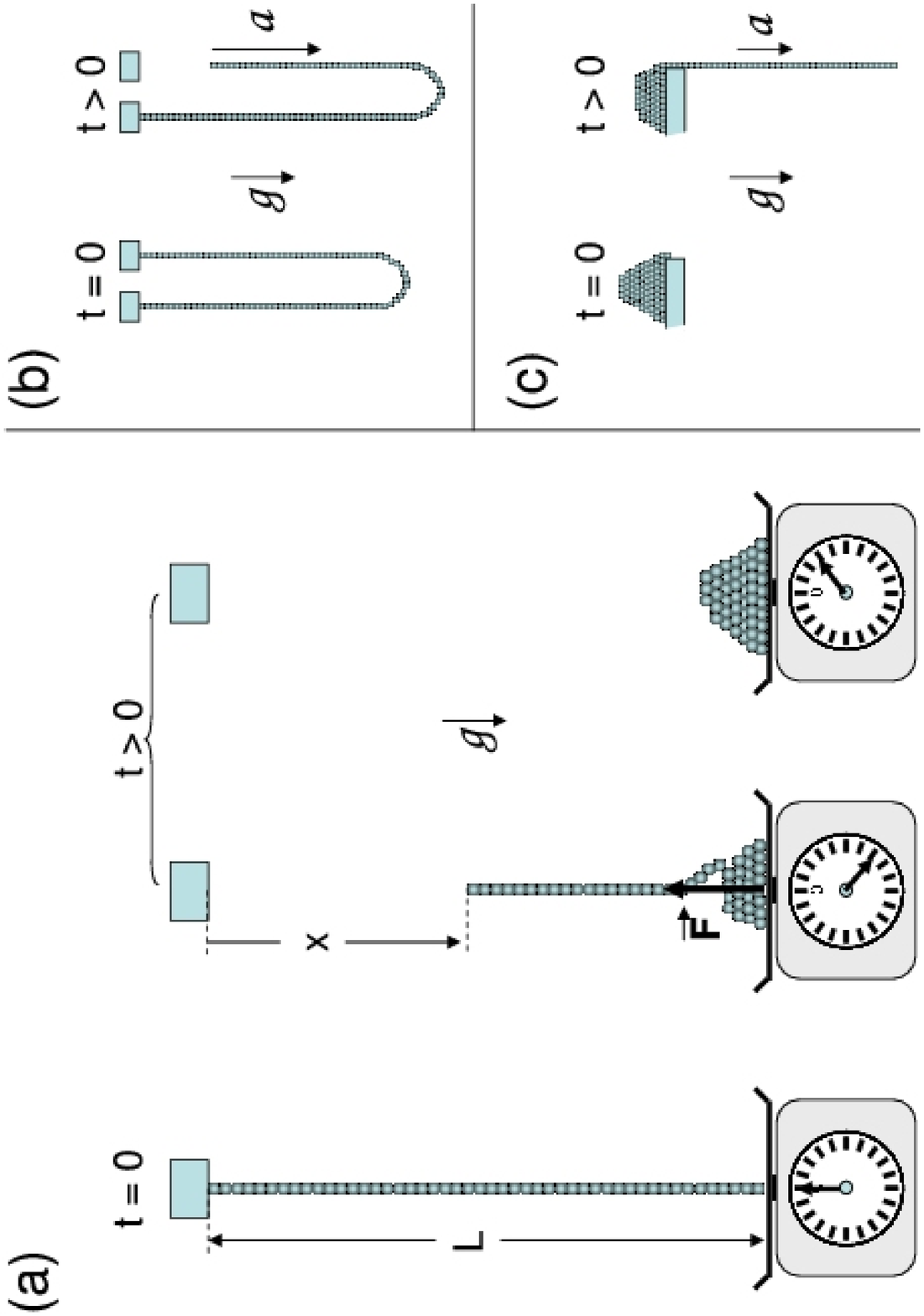}
\caption{Falling chain geometries. (a) The chain is falling vertically on a scale pan. During the fall, the apparent weight is predicted to be three times the deposited weight. (b) The chain is initially attached by its two ends at the same height. As one of the ends is released the acceleration of the chain tip is greater than $g$, the acceleration due to gravity. (c) The chain slides off a table by a vertically hanging end. In this case the acceleration is predicted to be $g/2$.}
\label{figure1}
\end{figure}

Hold one end of a chain above a scale so that the other end lightly touches the pan.
Now release it from rest [see Fig.~\ref{figure1}(a)].
Theory predicts that during the fall, the scale will display an increasing weight equal to three times the weight of the deposited part
of the chain and then a sudden drop to the chain weight once the chain lays completely
motionless on the pan.\cite{morin07}
This simple experiment, which can be performed with basic laboratory equipment, has
a few subtleties that make it worthwhile to do a thorough theoretical and experimental analysis.

This system belongs to the class of variable-mass problems usually considered in
undergraduate physics textbooks. Among the various possible experimental
configurations we cite a few that have been considered recent. For example, consider
a chain attached by its two ends at the same height and separated horizontally by a
certain distance [see Fig.~\ref{figure1}(b)]. As one of the ends is released the acceleration of the chain tip
is greater than $g$, the acceleration due to gravity.\cite{calkin89, tomaszewski06, geminard08}
Another configuration is a chain sliding off a table by a vertically hanging end
[see Fig.~\ref{figure1}(c)]. In this case the acceleration is only a fraction of $g$.\cite{chun07} Other configurations are discussed in Ref.~\onlinecite{morin07}.

In this paper we consider experimentally and theoretically the problem of a vertically falling chain
that impinges on the pan of a scale.
To our knowledge, the only experimental realization of this system has been reported by
van den Berg.\cite{vandenBerg98}
A theoretical description of the idealized problem, extended to the more general case
of an additional constant vertical force acting on the upper end, was performed by \v{S}\'ima and J. Podolsk\'y,\cite{sima05}
which also cites unavailable Czech and Russian bibliography,
especially the role of Graf Georg von Buquoy, who was
``the first who investigated mechanical systems with a varying mass.''
There is a mention in the solutions manual by Thornton and Marion\cite{thornton} of the possibility of the free upper end falling faster than a freely falling mass, which is a puzzling effect. The authors report a personal communication from M.\ G.\ Calkin who ``\ldots\ has found that experimentally the time of fall for this problem is consistently less $\ldots$ by about 1.5\%'' and ``\ldots\ also finds evidence that suggests the free fall treatment is more valid if the table is energy absorbing.'' We shall see that this effect is striking.

\section{\label{sec1}Theoretical background}

The dynamics of the falling chain is usually treated as follows. Consider an inextensible and perfectly flexible chain of length $L$
with a homogeneous linear mass density $\mu$.
At $t=0$ the lower end of the chain just touches the scale pan while the
upper end is released from rest [see Fig.~\ref{figure1}(a)].

For $t>0$ the free end of the chain has fallen a distance $x=x(t)$
so that a part of the weight $\vec{w}(t) = \mu x \vec{g}$
($\vec{g}$ is the acceleration due to gravity) lies at rest on the pan.
At the same time, the falling part of the chain (length $L-x$) moves downward with
velocity $\vec{v}$ (the chain is inextensible).
The total force, $\vec{F}$, exerted by the scale on
the chain at time $t$ opposes the weight $\vec{w}$ of the already deposited part of the chain
and exerts the force $\vec{f}$ necessary to stop the chain
links that collide with the pan.
The latter contribution can be estimated by considering
that during the time interval $\Delta t$, a mass $\mu v \Delta t$ of the chain ($v=|\vec{v}|$) with velocity $\vec{v}$ at the beginning of the interval collides with the pan and comes to rest
(assuming that the collision is perfectly inelastic).
The change in the linear vertical momentum is thus $\Delta \vec{p} = - \mu v \Delta t\,\vec{v}$. From Newton's second law, $\vec{f} = d\vec{p}/dt = - (\mu v)\vec{v}$.
The reading of the scale corresponds to the norm of $\vec{F}$:
\begin{equation}
W \equiv |\vec{F}| = \mu g x + \mu v^2.
\label{weight}
\end{equation}
This result holds if $\mu$ is constant and the links that collide with the pan stop.

It is usually further assumed that the chain experiences free fall.
In this case, $x=g t^2/2$, $v = \dot{x} = g t$, and we obtain [xx usually we write $y$ for the vertical displacement xx]
\begin{equation}
W = \frac{3}{2} \mu g^2 t^2 = 3 w(t).
\label{weightFF}
\end{equation}
During the fall, the apparent weight of the chain, $W$, increases quadratically in time and equals,
at each instant, three times the weight $w$ of the part of the chain already deposited on the scale pan. The duration of the free-fall over the length $L$ equals $T=\sqrt{2L/g}$.
At $t=T$, the last link touches the pan and a maximum value of $3 M g$ is attained ($M$ denotes the mass of the entire chain).
At this time the apparent weight $W$ instantaneously drops to $M g$, the weight of the chain
which sits at rest on the scale pan.

In this treatment the chain is assumed to be homogeneous (constant mass density $\mu$ along its length)
and has no limiting radius of curvature (it can be bent infinitely). In a real experiment these two assumptions are not fully satisfied. In particular, we shall see that the assumption of free-fall does not hold when the minimum radius of curvature is finite.

\section{\label{sec2}Setup and protocol}

\begin{figure}[h!]
\includegraphics[width=0.45\columnwidth]{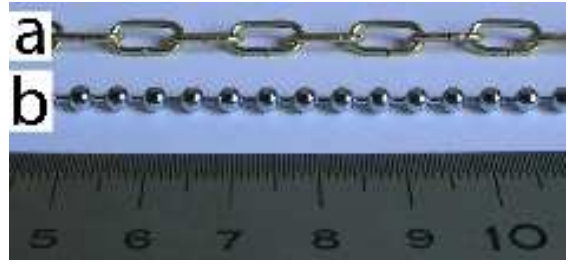}
\caption{Chains used in the experiments: {(a)} loop chain and {(b)} ball chain. The numbers on the ruler at the bottom are centimeters.}
\label{figure2}
\end{figure}

We shall report measurements obtained with two chains, namely a ``ball chain'' and a ``loop chain''
(see Fig.~\ref{figure2}).
The ball chain consists of stainless-steel identical links (length $\ell = (4.46 \pm 0.01)\,10^{-3}$\,m) [xx better to use $\ell$ since $l$ is difficult to distinguish from 1 xx]
that are each made from one rod and a hollow sphere (diameter $\phi = (3.26 \pm 0.01) \times 10^{-3}$\,m)
attached to each other.
The total length of the chain is $L=1.898 \pm 0.001$\,m, the number of links is $N = 413$,
and the total mass $M= (29.6 \pm 0.1)\times 10^{-3}$\,kg. The mass of one link is thus $m=M/N \approx 7 \times 10^{-5}$\,kg.
The loop chain is made of equal concatenated [xx what do you mean by concatenated ? xx] links (length $\ell = (10.22 \pm 0.01)\times10^{-3}$\,m,
width $(5.50 \pm 0.01)\times10^{-3}$\,m, thickness $(1.15 \pm 0.01)\times 10^{-3}$\,m).
Its total length, number of links, and mass are respectively $L=1.890 \pm 0.001$\,m, $N = 235$, and
$M= (41.8 \pm 0.1)\times 10^{-3}$\,kg.

\begin{figure}[h!]
\includegraphics[width=0.8\columnwidth]{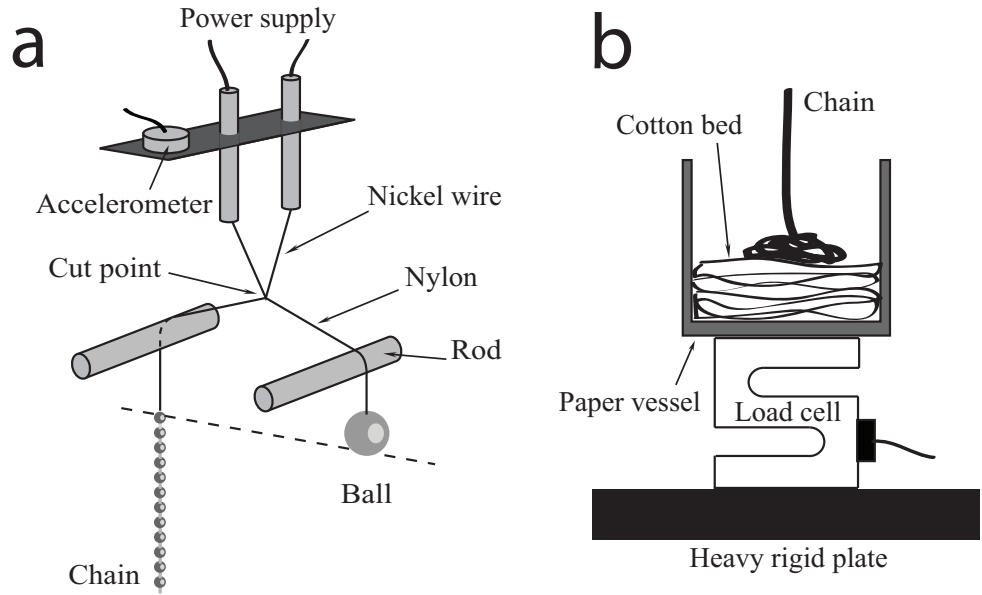}
\caption{Sketch of the experimental setup. (a) System used for the simultaneous release of the chain and of the steel ball. The dashed line illustrates that both the chain and ball are released from the same height. Alternatively the end of the nylon cord that holds the ball can be knotted to the right rod when the ball is not used. (b) Force measurement. At the bottom, the force applied by the falling chain on a vessel into which it falls is measured by a load cell.}
\label{figure3}
\end{figure}

The experimental setup (see Fig.~\ref{figure3}) consists of two parts.
The part at the top holds the chain previous to its release and
the part at the bottom measures the force exerted during its fall.

The system at the top [see Fig.~\ref{figure3}(a)] makes it possible to
simultaneously release the chain and a steel ball, so we can compare the motion of the chain with free-fall, and accurately determine the beginning of its motion.
It is an adaptation of the system described in Ref.~\onlinecite{geminard08}.
The upper end of the chain is initially attached with a nylon cord (fishing line, diameter 10$^{-4}$\,m)
to a steel ball such that the joining cord is held by a nickel wire (diameter 10$^{-4}$\,m).
Two horizontal parallel rods are used to separate the falling paths of the two bodies.
The whole system is adjusted so that initially the chain and the ball are at
the same height.
To initiate the fall, we quickly cut the nylon wire by injecting a large electric current
(about 1\,A) from a power supply in the metallic wire, during a few tenths of second.
The mount holding the nickel wire is a slightly-flexible cantilever spring, and the release of
the bodies results in a sudden upward acceleration of its free end. Measuring this acceleration
by means of an accelerometer (Dytran, 3035BG) makes it possible to accurately determine the origin of time, $t=0$.

The vertical component of the force applied by the falling chain
(apparent weight $W$) is measured by a force-sensor (Load cell, Testwell, KD40S, maximum force 10\,N) which holds a vessel made from a horizontal disk (Dural, diameter 6\,cm) surrounded by
high walls which prevent the chain or the steel ball from escaping the system
[see Fig.~\ref{figure3}(b)]. As described in the following, the vessel can be filled with
different materials to change the experimental conditions at the bottom. The material of the walls (paper or plastic) can also be changed.

The signals from the accelerometer and the force sensor are
monitored using an acquisition board (National Instruments, PCI 6251). Given that the typical length
of the chains used in the experiments is approximately 2\,m, the duration of the free fall is consistently about 0.6\,s (with $g=9.81$\ m/s$^{2}$). We acquired the signals at 10\,kHz, which is sufficient to reveal the details of the force dynamics while avoiding the need to manipulate large data sets.

\begin{figure}[h!]
\includegraphics[width=0.45\columnwidth]{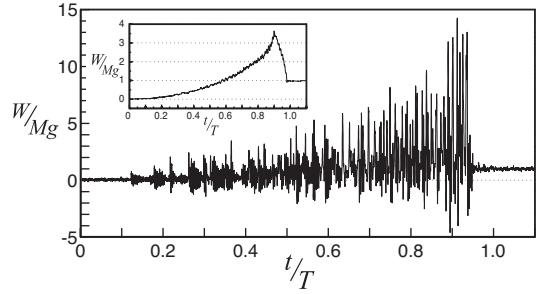}
\caption{The apparent weight $W$ versus the time $t$. $W$ is compared to the weight $M g$ of the chain, and the time $t$ is compared to the duration $T$ of free fall from the height $L$. The experiment is performed with the ball chain and the vessel is empty. Inset: filtered signal. The signal is smoothed by averaging over an interval of $\pm 50$\,ms. [xx isn't the interval necessarily positive? xx]}
\label{figure4}
\end{figure}

It is important to discuss the typical force-signal obtained
for a ball chain falling in an empty vessel (see Fig.~\ref{figure4}).
Instead of the continuous quadratic increase in the apparent weight $W$ with time,
we observe large fluctuations whose amplitude increases in time until the end of the fall.
The large peaks that characterize the force signal are a natural consequence of the discrete nature
of the chain.
The links hit the bottom of the vessel almost independently of each other and, as their velocity increases in time due to the acceleration of gravity, the amplitude
and the frequency of the peaks increase as well.
We could avoid the peaks by using a sensor with a long characteristic time.
However, the frequency of the collisions ranges from about 30 to 1300\,Hz during the fall.
The choice of a characteristic time larger than $1/30 \simeq 30$\,ms would hide some relevant details
of the fall. Thus, we chose a sensor with a resonant frequency
of nearly 500\,Hz, which makes it possible to obtain a temporal resolution of about 2\,ms. The ``hidden'' ideal force can then be revealed using a numerical filtering procedure
(see Fig.~\ref{figure4}, inset) that might wipe out relevant features of the signal. [xx I'm confused. if the force can be revealed, what is being wiped out? xx]

Instead of using a numerical filter, we look for a cleaner force signal
by changing the experimental conditions at the bottom.
To efficiently reduce the amplitude of the oscillations,
we filled the vessel successively with various viscous and non-newtonian fluids,
a granular material, diluted modeling clay, and cotton wool.
The force always exhibited violent oscillations except when a thick layer of cotton wool
(typically a few centimeters) was placed at the bottom of the vessel [see Fig.~\ref{figure3}(b)].
The signal from the force sensor obtained after addition of $1.5$\,g of cotton shows a monotonic increase in the force during the fall and a final drop
to the weight of the entire chain when its motion ceases in the vessel (seeFig.~\ref{figure5}).
The low-frequency oscillations (at $\sim 15$\,Hz) at the tail of the signal are produced
by restoring forces in the cotton bed. Our results show that the cotton layer constitutes a very efficient method of damping the force signal, and we therefore used it in our measurements.
One advantage is that we are able to identify short-time features of the force signal, in particular, the
time at which the last link of the chain hits the pan (maximum of the force), which otherwise would
be smoothed out by a numerical filtering procedure.

\section{\label{sec4} Results and discussion}

\begin{figure}[h!]
\includegraphics[width=0.45\columnwidth]{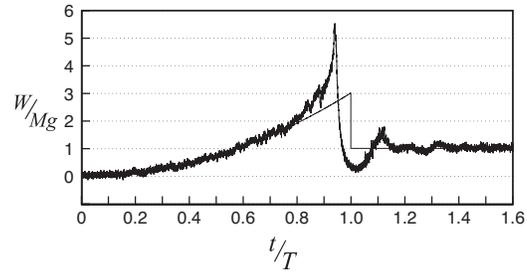}
\caption{Apparent weight $W$ versus time $t$. $W$ is compared to the weight $M g$ of the chain, and the time $t$ to the duration $T$ of free fall from the height $L$. The experiment is performed with the ball chain and with a cotton bed in the vessel. The amplitude of oscillations in the sensing system is notably reduced. The continuous line is the predicted apparent weight for an ideal free-falling chain.}
\label{figure5}
\end{figure}

We see in Fig.~\ref{figure5} that the
measured apparent weight does not follow the theoretical curve and, in particular,
the maximum force is much larger, about 5.5 times the weight of the chain.
We might claim that the difference originates from the condition at the bottom but,
even more striking than the value of the maximum force, we notice in Figs.~\ref{figure4}
and \ref{figure5} that the force drops to its rest value earlier than predicted for free fall. Whatever the condition at the bottom, the free end of the chain falls faster than expected.
The difference is significant, about $38$\,ms, corresponding to $12\%$ of the total length of the chain.

\begin{figure}[h!]
\includegraphics[width=0.45\columnwidth]{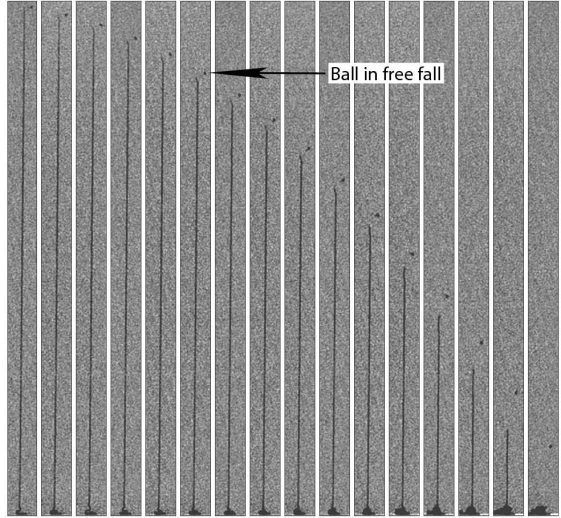}
\caption{Snapshots of the falling ball-chain with simultaneous drop of a steel ball. The free end of the chain (dark vertical line) falls faster than the steel ball (dark point on the right-hand side). The chain falls directly on a flat surface; note the formation of a compact heap at the bottom. Images are taken at intervals of $40\,\mu$s.}
\label{figure6}
\end{figure}

To check that the ball chain does fall faster than a simple body in free fall,
we photographed the simultaneous
falls of the chain and of the steel ball.
In this experiment the surface at the bottom is hard and flat (wood board
covered by an acrylic layer) so that the dissipation is much less than for the vessel padded with cotton wool. The sequence in Fig.~\ref{figure6} is conclusive:
the last link is clearly ahead of the ball during the entire fall, which
is possible only if, in addition to its own weight, a tensional force $\Gamma$
pulls the falling part of the chain downward. The origin of this force is surprising because the force exerted on the chain due to the pan is oriented upward.

It is interesting to consider several mechanisms that could alter the dynamics
of the chain and explain the deviation from free fall.

\textit{Air friction}. The falling chain is subjected to friction with the surrounding air.
This friction is the same along the chain, and thus does not generate any tension,
but rather reduces homogeneously the chain velocity.
This drag, which increases the time of fall,
is not responsible for the experimental observations.

\textit{Initial tension of the chain}. Previous to its release, the chain, which is hanging from its upper end,
is subjected to a tension $\Gamma$ as a consequence of its own weight.
Mechanical equilibrium imposes $\Gamma(s) = \mu g (L+s)$,
where $s$ denotes the position in a frame of coordinates attached to the chain,
oriented upward; the origin is attached to the uppermost link.
When the chain is released, a longitudinal wave propagates from top to bottom,
accelerating the links downward. From the value of the elastic constant associated with the stretching, $k = (5.5\pm0.4)\times 10^3$\,N
(defined by $\Gamma = k \Delta L/L$, where $\Delta L$
is the elongation and $L$ the initial length, and $\Gamma$ is obtained by measuring the elongation
due to known masses),
we estimate that the wave propagates at the velocity
$c_{l} = \sqrt{k/\mu} \approx 600$\,m/s, assuming the chain to be continuous.
Thus, the wave front reaches the lowermost link 3\,ms after the release of the uppermost one.
This relaxation leads to a contraction of about $MgL/(2k) \approx 50\,\mu$m of the chain
and to an acceleration of the center of mass toward the bottom.
It is interesting to consider the associated decrease in the fall time,
which can be estimated by an energy argument.
The total elastic energy initially stored
in the chain is $U = \frac{1}{6} (Mg)^2 (L/k)$.
If all the elastic energy was transferred to kinetic energy of the whole chain,
the latter would gain an initial velocity $v_0 = g L / (\sqrt{3}c_l) \approx 18 \times 10^{-3}$\,m/s,
derived from the condition $U = \frac{1}{2} M v_0^2$, which leads to a decrease in the fall time, $v_0/g$,
of about 2\,ms, much smaller than the experimental value of $38$\,ms (see Sec.~\ref{sec2}).
Therefore the initial tension of the chain does not explain the observed behavior, even for our
unrealistic hypothesis that all the elastic energy goes into kinetic energy.
In a process more consistent with momentum conservation the chain would rapidly dissipate
the elastic energy because of the friction associated with the relative motion of the links.
We can thus assume that the chain as a whole starts falling from rest at $t=0$,
in spite of its initial tension.

\textit{Finite bending limit of the chain}. We have to abandon the infinite flexibility approximation for a real chain, which can fold only over a finite distance with a geometry that depends on its construction. Specifically, there exists a minimum accessible radius of curvature, equal to a few times the link length. We shall see that this fact is responsible for the deviation of the experiments from the simplistic theory assuming free-fall.

\section{\label{sec5}Theoretical analysis}

What is the origin of the downward pulling force responsible for a net acceleration
larger than $g$, explaining the earlier arrival of the free end? A similar phenomenon is well known in the context of another variable-mass problem,
namely the partial fall of a chain that is held by its two ends at the same height
and then released at one of its ends.\cite{calkin89, tomaszewski06}
The free end is observed to fall faster than a simple body in free-fall.
This behavior is easy to understand if we consider that the energy is conserved during
the fall,\cite{wong06}
so that while more and more links come to rest, the potential energy is converted into
the kinetic energy of the falling part, which contains fewer and and links. The continuous decrease in the mass of the moving part results in a continuous increase
in the acceleration.
The energy is likely not to be conserved, due to internal dissipation associated with the relative
motion of the links.
Morin\cite{morin07} draws a distinction between a dissipative (loosely spaced links joined
by a massless string) and a non-dissipative (closely spaced links) chain, and discusses the influence of the folding on the dynamics for each case.

In our experimental configuration the energy is not conserved because
the chain remains at rest after the fall.
Therefore we will discuss a model based on Newton's laws.
We will see that the geometry of the folding at the bottom is crucial and
that, to account for the dynamics of the chain, we must
consider not only the momentum but also the angular momentum [xx you used kinetic moment, but that term is never used in any classical mechanics book I know of. I assume you mean angular momentum xx] of the system.
We will not solve a set of coupled equations modeling
the dynamics of each link. Such models exist,\cite{schagerl97, tomaszewski06}
but they would obscure the simple physics that is sufficient to understand the phenomenon.
We thus consider a minimal model containing the basic elements that describe
our observations.

\begin{figure}[h!]
\includegraphics[height=0.45\columnwidth]{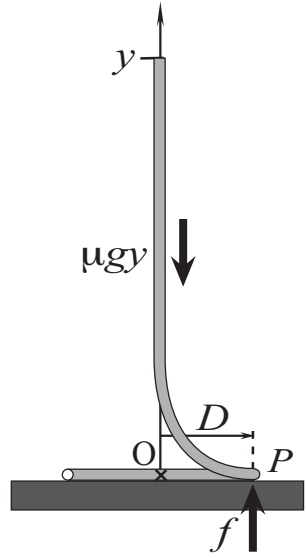}
\caption{\noindent Geometry of the falling chain in the simple model.}
\label{figure7}
\end{figure}

The main difference between the ideal case and the experiment is that the chain cannot be bent infinitely and, thus, exhibits a minimum accessible radius of curvature $R$.
The ideal chain would fall vertically toward a single point $O$.
Instead the real chain is straight and falls vertically in a large part of its length, but bends and curls close to the bottom in a region of typical size $\sim R$.
Based on the experimental observations (see Fig.~\ref{figure6}),
we approximate the shape of the falling chain according to the illustration in Fig.~\ref{figure7}.
We discard the details of the heap formation and the exact shape of the curled part, and
consider that the chain hits the bottom at the point P
which is a finite distance $D$ (of the order of $R$) from O.
The dynamical part $\vec{f}$ of the total force
exerted by the bottom on the chain is applied at P.

We denote by $y$, the vertical coordinate (oriented upward) of the upper free end. [xx here you use $y$, not $x$ xx]
We assume that $y \gg R$; that is, the length of the falling part
is much larger than the typical size of the curl. In this limit, we can neglect
the momentum of the curl with respect to that of the vertical part and write the equation
governing the change of momentum as
\begin{equation}
\frac{d}{dt}\Big(\mu y\dot{{y}}\Big) = -\mu y {g} + {f},
\label{momEq}
\end{equation}
where $\mu y$ corresponds to the falling mass and ${\dot y}$ to its vertical velocity.

In the same approximation, disregarding the rotation of the point P around O in Fig.~\ref{figure7},
we can estimate the horizontal component of the angular momentum, $\ang_{\rm O}$, of the chain about O to which only the
curl contributes, because the vertical part points toward O and the remaining part of the chain is at rest.
If we assume that in the curl, the links move with a velocity of the order of $\dot{y}$ at a typical
distance $R$ from O, we obtain $\ang_{\rm O} \sim \mu R^2 \dot{y}$
(where the sign is associated with a contact point on the right-hand side as sketched in
Fig.~\ref{figure7}). The change in the angular momentum is due to two contributions.
There is a change in the overall velocity of the chain $\dot{y}$, which leads to a term $\mu R^2 \ddot{y}$.
Also the links in the curl come to rest. In the curl the angular momentum goes from
$\sigma_{\rm O}$ to $0$ in the typical time $\tau \simeq -R/\dot{y}$ ($\dot{y} < 0$). This contribution thus is estimated to be of the order of $-\ang_{\rm O}/\tau \simeq \mu R \dot{y}^2$.
According to Newton's laws, we obtain
\begin{equation}
\frac{d\ang_{\rm O}}{dt} = D f \simeq \mu (R^2 \ddot{y} + R \dot{y}^2).
\label{angMomEq}
\end{equation}

We combine Eqs.~(\ref{momEq}) and (\ref{angMomEq}) and assume that $R \ll y$
[which leads to neglecting the term $\mu R^2 \ddot{y}$ in Eq.~(\ref{angMomEq})] and
obtain the differential equation for $y(t)$:
\begin{equation}
\ddot{y}=-g+(\gamma-1) \frac{\dot{y}^2}{y},
\label{tipEqDim}
\end{equation}
where the geometrical factor $\gamma$ accounts for the shape of the curl.
Equation~(\ref{angMomEq}), with the same approximation, gives the dynamical force
\begin{equation}
f=\gamma \mu \dot{y}^2.
\label{forceDim}
\end{equation}
Accordingly, the apparent weight measured at the bottom is
\begin{equation}
W = \mu (L-y)g+\gamma \mu \dot{y}^2.
\label{AppW}
\end{equation}

If we write $x=L-y$, we recover Eq.~(\ref{weight}) for $\gamma=1$,
the case of an ideal free-falling chain.
For $\gamma \neq 1$, Eq.~(\ref{tipEqDim}) must be solved numerically
for the initial conditions $y(0)=L$ and $\dot{y}(0)=0$.
It is useful to define the dimensionless quantities $\tau \equiv t/T$, $Y \equiv y/L$,
and $\Omega \equiv W/(M g)$, where $T \equiv \sqrt{2L/g}$ is the time of free-fall from the height $L$.
Equation~(\ref{tipEqDim}) becomes
\begin{equation}
\ddot{Y} = -2+(\gamma-1)\frac{\dot{Y}^2}{Y},
\label{tipEqAdim}
\end{equation}
and must be solved with the initial conditions $Y(0)=1$ and $\dot{Y}(0)=0$. The apparent weight is given by
\begin{equation}
\Phi=1-Y+\frac{\gamma}{2} \dot{Y}^2.
\label{totalForceAdim}
\end{equation}

\begin{figure}[h!]
\includegraphics[width=0.45\columnwidth]{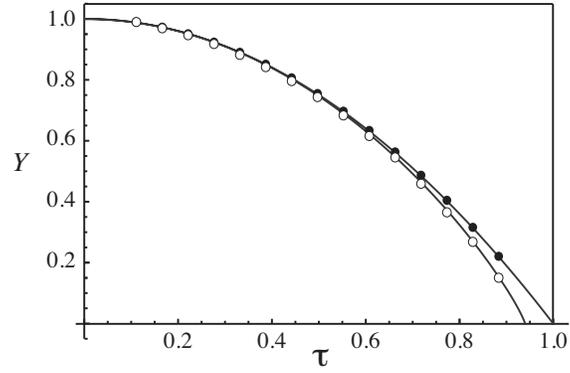}
\caption{\label{figure8}The vertical position $y$ [xx note change xx] versus the time $\tau$. In the experiment the ball chain (open circles) and the steel ball (full circles) fall simultaneously (data obtained from Fig.~\ref{figure6}). The line joining the open circles corresponds to the integration of Eq.~(\ref{tipEqAdim}) with $\gamma = 0.83$.}
\end{figure}

The numerical solution of Eq.~(\ref{tipEqAdim}) best interpolates the measured values of the free-end position for $\gamma = 0.83$ (see Fig.~\ref{figure8}). The associated duration of the fall is $\tau \simeq 0.94$, which corresponds to the free end reaching the bottom $36$\,ms earlier than expected for free fall. This result compares to the experimental value of 38\,ms given in Sec.~\ref{sec4}.

\begin{figure}[h!]
\includegraphics[width=0.45\columnwidth]{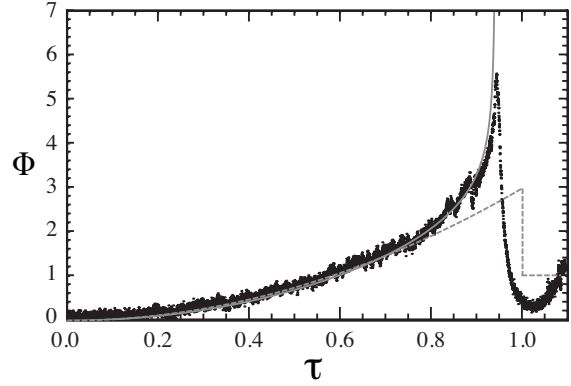}
\caption{\label{figure9}Experimental force (black points) compared to the prediction of the model (continuous line) with $\gamma = 0.83$. Also shown is the ideal force for a chain with the same length as the one used in the experiment (dashed line).}
\end{figure}

The apparent weight predicted by Eq.~(\ref{totalForceAdim}) for $\gamma = 0.83$ is compared with the experimental measurements in Fig.~\ref{figure9}.
For the same value of $\gamma$
we can interpolate the kinematic curve
(see Fig.~\ref{figure8}) obtained for the fall onto a hard surface
and the dynamic curve (Fig.~\ref{figure9}) obtained for the fall on a cotton wool layer.
Thus, $\gamma$ is independent of the dissipation at the bottom.
Note that the simple model fails to account for the finite value of the
force because the approximation $R \ll y$ is no longer satisfied when the free end reaches the bottom.

Let us determine the tension that acts, according to our model, on the falling part of the chain
and which is responsible for the larger than $g$ acceleration of the free end.
For a length element $ds$ of the chain and the tension $\Gamma$,
we can write the change in the momentum as
\begin{equation}
\mu ds \ddot{y} = - (\mu ds) g + \frac{\partial \Gamma}{\partial s} dl.
\end{equation}
Thus, at a given time $t$, the tension in the chain $\Gamma(s) = \mu (\ddot{y}+g) s$,
where the condition $\Gamma(0) = 0$ at the free end is already taken into account.
From Eq.~(\ref{tipEqDim}), we can determine the tension pulling the chain toward the bottom at $s =-y$:
\begin{equation}
\Gamma(-y) = \mu (1-\gamma) \dot{y}^2
\label{tension}
\end{equation}
The chain cannot transmit a compressive stress, which leads to the condition
that $\Gamma \ge 0$ and, thus, that $\gamma \le 1$. Note that, in the limit $\gamma=1$, we recover the ideal case in which no tension exists.
Also whatever the value of $\gamma$, the total force applied to the curl
equals $(\Gamma(-y)+f) = \mu \dot{y}^2$. Thus, in all cases
the total force applied to the curl is correctly predicted by considering
the change in the linear momentum of the links per unit time in that region [xx what region? xx],
as described in Sec.~\ref{sec1}.
During a time $\Delta t$ a length $-\dot{y}\, \Delta t$ changes from the vertical velocity $\dot{y}$ to rest,
so that the total vertical applied force must be
\begin{equation}
\mu (-\dot{y}\, \Delta t) \frac{(0-\dot{y})}{\Delta t} = \mu \dot{y}^2.
\end{equation}
The difference between the measured apparent weight and $\mu \dot{y}^2$
is thus due to the internal tension of the chain at the bottom.

\section{\label{sec6}Discussion and Conclusion}

We have demonstrated that the existence of a minimum radius of curvature,
which imposes a curl at the bottom, is responsible for the appearance of
an additional tension in a falling chain, which pulls the falling part toward the bottom.
However, the model discussed in Sec.~\ref{sec5} leaves a few open questions.

In accord with our observations, we assumed that the chain falls
vertically, except in a small region of size $R$ at the bottom.
Nonetheless, the deformation of the chain at the bottom excites transverse waves that can
eventually climb along the chain. No such waves were observed.
To understand this behavior, we consider the velocity, $c_t = \sqrt{\Gamma/\mu}$,
of the transverse waves associated with the tension $\Gamma$. At the bottom, where $\Gamma$ (and thus $c_t$)
is maximum, we have, from Eq.~(\ref{tension}), $c_t = \sqrt{1-\gamma} |\dot{y}| < |\dot{y}|$.
The velocity $c_t$, even at the bottom, is always smaller than the fall velocity
so that the perturbations at the bottom cannot propagate upward, which is the reason why the chain
remains straight and vertical and does not move laterally.

As mentioned, the energy is not conserved. The total energy of the
chain is $E = \mu y \dot{y}^2/2 + \mu y^2 g/2$.
If we differentiate $E$ with respect to time and use Eq.~(\ref{tipEqDim}), we
one obtain the rate of energy dissipation:
\begin{equation}
\frac{dE}{dt}= \left(\gamma - \frac{1}{2} \right) \mu \dot{y}^3.
\label{denergy}
\end{equation}
As a consequence, because $\dot{y}<0$, we must have $\gamma > 1/2$.
We know that $\gamma \le 1$ from the condition that the chain cannot transmit
a compressive stress.
Thus, $\gamma$ must be in the range $[1/2,1]$ and therefore the fall time ranges
from the time of free fall $T$ for $\gamma = 1$ to $0.847\,T$ for $\gamma = 1/2$,
in the limit where energy is conserved.
The experimental value $\gamma \simeq 0.83$ obtained for the ball chain is in $[1/2,1]$.

The dimensionless parameter $\gamma$ in our model t accounts for the geometry of the curl at the contact with the bottom. It might be tempting to interpret $\gamma$ as characterizing the dissipation at the bottom. However, the values of $\gamma$ measured for a hard or a soft surface are the same, in contrast with Ref.~\onlinecite{thornton}, but in agreement with our theoretical interpretation. If $\gamma$ characterizes the geometry of the curl, it must depend on the structure of the chain. For instance, the ball chain exhibits a minimum radius of curvature of about $2$ links.
In contrast, the classical loop chain used in previous experiments\cite{vandenBerg98} does not exhibit a well-defined minimum radius of curvature and can apparently be bent without any resistance. In a similar experiment with the loop chain we obtain $\gamma \simeq 0.95$, again in the interval $[1/2,1]$, and we find that the free end reaches the bottom 8\,ms earlier than the body in free fall. For the loop chain, the force is applied at a point almost aligned with the vertical falling part (P is close to O in Fig.~\ref{figure7}), and the effect is weaker than for the ball chain.

Our model does not give an estimate
the maximum apparent weight. $W$ is predicted to diverge at the end of the fall, when $y$ vanishes, which is not physically acceptable. The discrepancy between the model and the experiment is not surprising in this limit, because the assumption $R \ll y$ does not hold at the end of the fall. The analysis developed in Sec.~\ref{sec5} can be refined to include this effect, but we would not learn much more. Our analysis correctly describes the temporal behavior of the apparent weight during most of the fall and explains why the free end reaches the bottom earlier than a mass in free fall.

\begin{acknowledgments}
The authors would like to thank R.\ Freund for introducing them to the problem,
and H.\ Gayvallet and B.\ Castaing for fruitful discussions.
E.\ H.\ thanks the ENS de Lyon for having invited him as a professor in July 2009, during which this work was done.
\end{acknowledgments}

\section*{Figure captions}

\newpage

\newpage

\newpage

\newpage

\newpage

\newpage

\newpage

\newpage

\newpage

\end{document}